\documentstyle[11pt]{article}
\begin{document}

\title{Decomposition of higher  order equations of Monge-Ampere type}

\author{{\Large Ferapontov E.V. } \\
    Department of Mathematical Sciences \\
    Loughborough University \\
    Loughborough, Leicestershire LE11 3TU \\
    United Kingdom \\
    e-mail: 
    {\tt E.V.Ferapontov@lboro.ac.uk}
}
\date{}
\maketitle

\newtheorem{theorem}{Theorem}
\newtheorem{proposition}{Proposition}
\newtheorem{lemma}{Lemma}

\pagestyle{plain}

\maketitle

\begin{abstract}

It is demonstrated that the fourth order PDE
$$
{\rm det} \left|
\begin{array}{ccc}
f_{xxxx} & f_{xxxt} & f_{xxtt} \\
f_{xxxt} & f_{xxtt} & f_{xttt} \\
f_{xxtt} & f_{xttt} & f_{tttt}
\end{array}
\right| =0
$$
decouples in a pair of identical second order Monge-Ampere equations,
$$
u_{xx}u_{tt}-u_{xt}^2=0 ~~ {\rm and} ~~ v_{xx}v_{tt}-v_{xt}^2=0,
$$
by virtue of the B\"acklund-type relation
$$
d^4f=\frac{(d^2u)^2}{u_{xx}} + \frac{(d^2v)^2}{v_{xx}}.
$$
A higher order generalization of this decomposition ia also proposed.
\bigskip



\end{abstract}

\section{Introduction}
Given a function $f(x, t)$, its fourth differential,
$$
d^4f=f_{xxxx}\ dx^4+4f_{xxxt}\ dx^3dt+6f_{xxtt}\ dx^2dt^2+4f_{xttt}\
dxdt^3+f_{tttt}\ dt^4,
$$
is a quartic form in $dx, dt$. It is well-known that any quartic form in two
variables can be expressed as a
sum of three 4th powers of linear forms,
see e.g. \cite{Harris}, Corollary 11.34. The particular case of  two 4th
powers is characterised by
the vanishing of the catalecticant determinant
\begin{equation}
{\rm det} \left|
\begin{array}{ccc}
f_{xxxx} & f_{xxxt} & f_{xxtt} \\
f_{xxxt} & f_{xxtt} & f_{xttt} \\
f_{xxtt} & f_{xttt} & f_{tttt}
\end{array}
\right| =0,
\label{cat}
\end{equation}
which is a fourth order PDE for the function $f$. Equivalently, condition
(\ref{cat}) implies the
harmonicity of the four zero directions of the differential $d^4f$, that is,
the cross-ratio thereof is equal to $-1$.
Equation (\ref{cat}) is  completely exceptional
in the sense of \cite{Boillat}. We recall that a generic completely
exceptional fourth order PDE
of  Monge-Ampere type can be defined by
equating to zero a linear combination of minors (of all possible orders  1,
2, and 3) of the 
catalecticant matrix  (\ref{cat}). All these equations are known to be
linearly degenerate
and satisfy a number of remarkable geometric properties.

For a function $f$ solving the equation (\ref{cat}),
its fourth differential is a sum of two 4th powers of linear forms,
\begin{equation}
d^4f=\mu_1(dx+\lambda_1dt)^4 + \mu_2(dx+\lambda_2dt)^4,
\label{1}
\end{equation}
or, explicitly,
\begin{equation}
\begin{array}{c}
f_{xxxx}=\mu_1+\mu_2, \\
\ \\
f_{xxxt}=\mu_1\lambda_1+\mu_2\lambda_2, \\
\ \\
f_{xxtt}=\mu_1\lambda_1^2+\mu_2\lambda_2^2, \\
\ \\
f_{xttt}=\mu_1\lambda_1^3+\mu_2\lambda_2^3, \\
\ \\
f_{tttt}=\mu_1\lambda_1^4+\mu_2\lambda_2^4.
\end{array}
\label{2}
\end{equation}

\bigskip

{\bf Remark.} It should be pointed out that $\mu_i$ and $\lambda_i$ may  be
complex. For instance,
the fourth differential of the function $f=\frac{1}{6}(x^3t-t^3x)$, which
clearly solves (\ref{cat}), is
$$
d^4f=dx^3dt-dt^3dx=dx\ dt\ (dx-dt)(dx+dt)=\prod _1^4(dx-c_idt),
$$
where  $(c_1, c_2, c_3, c_4)=(0, \infty, 1, -1)$. The corresponding
cross-ratio is
$$
\frac{(c_1-c_3)(c_2-c_4)}{(c_2-c_3)(c_1-c_4)}=-1,
$$
so that  zero directions of $d^4f$ are indeed harmonic. However, the
representation (\ref{1}) is complex,
$d^4f=((dx+idt)^4-(dx-idt)^4)/8i.$

\bigskip

Compatibility conditions of  equations (\ref{2}) imply that both
$(\lambda_1, \mu_1)$ and $(\lambda_2, \mu_2)$
solve one and the same system
$$
\lambda_t=\lambda \lambda_x, ~~ \mu_t=(\lambda \mu)_x
$$
which is equivalent to the Monge-Ampere equation
\begin{equation}
u_{xx}u_{tt}-u_{xt}^2=0
\label{Monge}
\end{equation}
under the transformation $u_{xx}=\mu, \ u_{xt}=\lambda \mu, \
u_{tt}=\lambda^2 \mu$ \cite{Nutku}.
Thus, equation (\ref{cat})
is a direct sum of two identical copies of the Monge-Ampere equation
(\ref{Monge})! A higher order version of this
decomposition is discussed in sect. 2. Choosing two particular solutions of
(\ref{Monge}), say $u(x, t)$ and $v(x, t)$, and
setting
$$
\mu_1=u_{xx}, ~~ \mu_1\lambda_1=u_{xt}, ~~ \mu_1\lambda_1^2=u_{tt},
$$
$$
\mu_2=v_{xx}, ~~ \mu_2\lambda_2=v_{xt}, ~~ \mu_2\lambda_2^2=v_{tt},
$$
one readily rewrites (\ref{1}) in the form
$$
d^4f=\frac{(d^2u)^2}{u_{xx}} + \frac{(d^2v)^2}{v_{xx}}
$$
which, after a partial integration, simplifies to
$$
f_{xx}=u+v, ~~~ f_{xttt}=\frac{u_{tt}^2}{u_{xt}}+\frac{v_{tt}^2}{v_{xt}},
~~~ 
f_{tttt}=\frac{u_{tt}^2}{u_{xx}}+\frac{v_{tt}^2}{v_{xx}}.
$$
Notice that equations (\ref{2}) can be solved for $\mu$ and $\lambda$,
indeed, rewriting (\ref{2}) in matrix form,
$$
\left(
\begin{array}{ccc}
f_{xxxx} & f_{xxxt} & f_{xxtt} \\
f_{xxxt} & f_{xxtt} & f_{xttt} \\
f_{xxtt} & f_{xttt} & f_{tttt}
\end{array}
\right) =\mu_1
\left(
\begin{array}{ccc}
1 & \lambda_1 & \lambda_1^2 \\
\lambda_1 & \lambda_1^2 & \lambda_1^3 \\
\lambda_1^2 & \lambda_1^3 & \lambda_1^4
\end{array}
\right)+
\mu_2
\left(
\begin{array}{ccc}
1 & \lambda_2 & \lambda_2^2 \\
\lambda_2 & \lambda_2^2 & \lambda_2^3 \\
\lambda_2^2 & \lambda_2^3 & \lambda_2^4
\end{array}
\right),
$$
one obtains that both $(\lambda_1, \mu_1)$ and  $(\lambda_2, \mu_2)$ must
satisfy the condition
\begin{equation}
{\rm rank}\left(
\left(
\begin{array}{ccc}
f_{xxxx} & f_{xxxt} & f_{xxtt} \\
f_{xxxt} & f_{xxtt} & f_{xttt} \\
f_{xxtt} & f_{xttt} & f_{tttt}
\end{array}
\right) -\mu
\left(
\begin{array}{ccc}
1 & \lambda & \lambda^2 \\
\lambda & \lambda^2 & \lambda^3 \\
\lambda^2 & \lambda^3 & \lambda^4
\end{array}
\right)\right)=1.
\label{lm}
\end{equation}
This condition clearly implies that $\lambda_1$ and $\lambda_2$  are two
roots of the quadratic equation
$$
{\rm det} \left|
\begin{array}{ccc}
f_{xxxx} & f_{xxxt} & f_{xxtt} \\
f_{xxxt} & f_{xxtt} &  f_{xttt} \\
1 & \lambda & \lambda^2
\end{array}
\right| =0,
$$
(or any of the equivalent equations obtained by replacing any other row of
the matrix (\ref{cat}) by the row
$(1, \ \lambda, \ \lambda^2)$). The corresponding $\mu$'s can be obtained by
equating to zero, say, the first  $2\times 2$
minor of the matrix (\ref{lm}), which gives
$$
\mu=\frac{f_{xxxx}f_{xxtt}-f_{xxxt}^2}{f_{xxxx}\lambda^2-2\lambda
f_{xxxt}+f_{xxtt}}.
$$

{\bf Remark.} Introducing, for simplicity, the notation
$$
f_{xxxx}=a, \  f_{xxxt}=b, \   f_{xxtt}=c, \  f_{xttt}=d, \  f_{tttt}=e,
$$
we readily see that equation (\ref{cat}) defines a cubic hypersurface in the
projective space $P^4$ with homogeneous coordinates
$(a : b : c : d : e)$. The change of variables (\ref{2}) has a simple
algebro-geometric interpretation, expressing the
well-known fact that this cubic is a bisecant variety of the norm-curve
$\gamma=(1 : \lambda : \lambda^2 : \lambda^3 : \lambda^4)$,
that is, the variety spanned by all lines intersecting $\gamma$ at two
points, see e.g. \cite{Harris},
Proposition 9.7. 

The comparison of Lie-point symmetries of equations (\ref{Monge}) and
(\ref{cat}) demonstrates that
(\ref{cat}) does not inherit the projective invariance of (\ref{Monge})
(sect. 3).

\section{Higher order generalizations}

For simplicity,  we consider  the $4\times 4$ catalecticant determinant
\begin{equation}
{\rm det} \left|
\begin{array}{cccc}
f_{xxxxxx} & f_{xxxxxt} & f_{xxxxtt}& f_{xxxttt} \\
f_{xxxxxt} & f_{xxxxtt} & f_{xxxttt}& f_{xxtttt} \\
f_{xxxxtt} & f_{xxxttt} & f_{xxtttt}& f_{xttttt} \\
f_{xxxttt} & f_{xxtttt} & f_{xttttt}& f_{tttttt} \\
\end{array}
\right| =0,
\label{cat1}
\end{equation}
which is a 6th order PDE in $f$. The vanishing of this determinant implies
that the differential $d^6f$ is a sum of three
6th powers of linear forms (one needs four 6th powers for a generic sextic),
\begin{equation}
d^6f=\mu_1(dx+\lambda_1dt)^6 + \mu_2(dx+\lambda_2dt)^6
+\mu_3(dx+\lambda_3dt)^6,
\label{*}
\end{equation}
or, explicitly,
\begin{equation}
\begin{array}{c}
f_{xxxxxx}=\mu_1+\mu_2+\mu_3, \\
\ \\
f_{xxxxxt}=\mu_1\lambda_1+\mu_2\lambda_2 +\mu_3\lambda_3, \\
\ \\
f_{xxxxtt}=\mu_1\lambda_1^2+\mu_2\lambda_2^2+ \mu_3\lambda_3^2, \\
\ \\
f_{xxxttt}=\mu_1\lambda_1^3+\mu_2\lambda_2^3 + \mu_3\lambda_3^3, \\
\ \\
f_{xxtttt}=\mu_1\lambda_1^4+\mu_2\lambda_2^4 + \mu_3\lambda_3^4, \\
\ \\
f_{xttttt}=\mu_1\lambda_1^5+\mu_2\lambda_2^5 + \mu_3\lambda_3^5, \\
\ \\
f_{tttttt}=\mu_1\lambda_1^6+\mu_2\lambda_2^6 + \mu_3\lambda_3^6. \\
\end{array}
\label{4}
\end{equation}
Geometrically, formulae (\ref{4}) manifest the fact that the quartic
hypersurface in $P^6$ defined by equation
(\ref{cat1}), is the trisecant variety of the norm-curve
$\gamma=(1: \lambda :  \lambda^2 : \lambda^3 : \lambda^4 : \lambda^5 :
\lambda^6)$, that is, the variety spanned by all
2-planes which intersect $\gamma$ at three points.

Again, compatibility conditions of  equations (\ref{4}) imply that each pair
$(\lambda_1, \mu_1)$,  $(\lambda_2, \mu_2)$
and $(\lambda_3, \mu_3)$ satisfies  the system
$$
\lambda_t=\lambda \lambda_x, ~~ \mu_t=(\lambda \mu)_x,
$$
so that equation (\ref{cat1})
is a direct sum of three identical copies of the Monge-Ampere equation
(\ref{Monge}). 
The reader may easily generalize this result to
arbitrary even
order. Choosing three particular solutions of (\ref{Monge}), say $u, v$ and
$w$, and
setting
$$
\mu_1=u_{xx}, ~~ \mu_1\lambda_1=u_{xt}, ~~ \mu_1\lambda_1^2=u_{tt},
$$
$$
\mu_2=v_{xx}, ~~ \mu_2\lambda_2=v_{xt}, ~~ \mu_2\lambda_2^2=v_{tt},
$$
$$
\mu_3=w_{xx}, ~~ \mu_3\lambda_3=w_{xt}, ~~ \mu_3\lambda_3^2=w_{tt},
$$
one readily rewrites (\ref{*}) in the form
$$
d^6f=\frac{(d^2u)^3}{u_{xx}^2} + \frac{(d^2v)^3}{v_{xx}^2}+
\frac{(d^2w)^3}{w_{xx}^2}.
$$
Notice that equations (\ref{4}) can be  written  in matrix form,
$$
\left(
\begin{array}{cccc}
f_{xxxxxx} & f_{xxxxxt} & f_{xxxxtt} & f_{xxxttt} \\
f_{xxxxxt} & f_{xxxxtt} & f_{xxxttt} & f_{xxtttt} \\
f_{xxxxtt} & f_{xxxttt} & f_{xxtttt} & f_{xttttt} \\
f_{xxxttt} & f_{xxtttt} & f_{xttttt} & f_{tttttt}
\end{array}
\right) 
=\sum_1^3
\mu_i
\left(
\begin{array}{cccc}
1 & \lambda_i & \lambda_i^2 & \lambda_i^3\\
\lambda_i & \lambda_i^2 & \lambda_i^3 & \lambda_i^4 \\
\lambda_i^2 & \lambda_i^3 & \lambda_i^4 & \lambda_i^5 \\
\lambda_i^3 & \lambda_i^4 & \lambda_i^5 & \lambda_i^6
\end{array}
\right),
$$
which implies that each pair $(\lambda_i, \mu_i)$ must satisfy the condition
\begin{equation}
{\rm rank}\left(
\left(
\begin{array}{cccc}
f_{xxxxxx} & f_{xxxxxt} & f_{xxxxtt} & f_{xxxttt} \\
f_{xxxxxt} & f_{xxxxtt} & f_{xxxttt} & f_{xxtttt} \\
f_{xxxxtt} & f_{xxxttt} & f_{xxtttt} & f_{xttttt} \\
f_{xxxttt} & f_{xxtttt} & f_{xttttt} & f_{tttttt}
\end{array}
\right) 
-\mu
\left(
\begin{array}{cccc}
1 & \lambda & \lambda^2 & \lambda^3\\
\lambda & \lambda^2 & \lambda^3 & \lambda^4 \\
\lambda^2 & \lambda^3 & \lambda^4 & \lambda^5 \\
\lambda^3 & \lambda^4 & \lambda^5 & \lambda^6
\end{array}
\right)
\right)=2.
\label{lm1}
\end{equation}
This condition clearly implies that  $\lambda$'s  solve the cubic equation
$$
{\rm det} \left|
\begin{array}{cccc}
f_{xxxxxx} & f_{xxxxxt} & f_{xxxxtt}& f_{xxxttt} \\
f_{xxxxxt} & f_{xxxxtt} & f_{xxxttt}& f_{xxtttt} \\
f_{xxxxtt} & f_{xxxttt} & f_{xxtttt}& f_{xttttt} \\
1 & \lambda & \lambda^2 & \lambda^3 \\
\end{array}
\right| =0,
$$
(or either of the equivalent equations obtained by replacing any other row
of the matrix (\ref{cat1}) by the row
$(1, \ \lambda, \ \lambda^2, \ \lambda^3)$). The corresponding $\mu$'s can
be obtained by equating to zero
any of the   $3\times 3$
minors of the matrix (\ref{lm1}).

\section{Point symmetries}

It is well-known, see e.g. \cite{Chupakhin},  that the algebra of Lie-point
symmetries of the 
Monge-Ampere equation (\ref{Monge})
is 15-dimensional, with infinitesimal generators
$$
\begin{array}{c}
{\bf X}_1=\frac{\partial}{\partial u}, ~~ {\bf X}_2=x
\frac{\partial}{\partial u}, ~~
{\bf X}_3=t\frac{\partial}{\partial u}, ~~ {\bf
X}_4=u\frac{\partial}{\partial u}, ~~
{\bf X}_5=\frac{\partial}{\partial x}, ~~ {\bf X}_6=\frac{\partial}{\partial
t}, \\
\ \\
{\bf X}_7=x\frac{\partial}{\partial x}, ~~ {\bf
X}_8=x\frac{\partial}{\partial t}, ~~
{\bf X}_9=t\frac{\partial}{\partial x}, ~~{\bf
X}_{10}=t\frac{\partial}{\partial t}, ~~
{\bf X}_{11}=u\frac{\partial}{\partial x}, ~~{\bf
X}_{12}=u\frac{\partial}{\partial t}, \\
\ \\
{\bf X}_{13}=ux\frac{\partial}{\partial x} + ut\frac{\partial}{\partial
t}+u^2\frac{\partial}{\partial u}, \\
\ \\
{\bf X}_{14}=x^2\frac{\partial}{\partial x} + xt\frac{\partial}{\partial
t}+ux\frac{\partial}{\partial u}, \\
\ \\
{\bf X}_{15}=xt\frac{\partial}{\partial x} + t^2\frac{\partial}{\partial
t}+ut\frac{\partial}{\partial u}.
\end{array}
$$
This algebra is isomorphic to  $SL(4)$, the Lie algebra of projective group
(indeed,  equation (\ref{Monge}) is descriptive of developable surfaces
which are an object of projective geometry).
Although  equation (\ref{cat}) is a direct sum of  two identical copies of
(\ref{Monge}), it does not inherit
the projective invariance of (\ref{Monge}):
the algebra of Lie-point symmetries of  (\ref{cat}) is 19-dimensional, with
infinitesimal generators
\begin{equation}
\begin{array}{c}
{\bf Y}_1=\frac{\partial}{\partial f}, ~~ {\bf Y}_2=x
\frac{\partial}{\partial f}, ~~
{\bf Y}_3=t\frac{\partial}{\partial f}, ~~ {\bf
Y}_4=x^2\frac{\partial}{\partial f}, ~~
{\bf Y}_5=xt\frac{\partial}{\partial f}, ~~{\bf
Y}_6=t^2\frac{\partial}{\partial f}, \\
\ \\
 {\bf Y}_7=x^3\frac{\partial}{\partial f}, ~~
{\bf Y}_8=x^2t\frac{\partial}{\partial f}, ~~
{\bf Y}_9=xt^2\frac{\partial}{\partial f}, ~~ {\bf
Y}_{10}=t^3\frac{\partial}{\partial f}, \\
\ \\
{\bf Y}_{11}=f\frac{\partial}{\partial f}, ~~
{\bf Y}_{12}=\frac{\partial}{\partial x}, ~~ {\bf
Y}_{13}=\frac{\partial}{\partial t}, ~~
{\bf Y}_{14}=x\frac{\partial}{\partial x}, \\
\ \\
{\bf Y}_{15}=x\frac{\partial}{\partial t}, ~~
{\bf Y}_{16}=t\frac{\partial}{\partial x}, ~~{\bf
Y}_{17}=t\frac{\partial}{\partial t}, \\
\ \\
{\bf Y}_{18}=x^2\frac{\partial}{\partial x} + xt\frac{\partial}{\partial
t}+3xf\frac{\partial}{\partial f}, \\
\ \\
{\bf Y}_{19}=xt\frac{\partial}{\partial x} + t^2\frac{\partial}{\partial
t}+3tf\frac{\partial}{\partial f}.
\end{array}
\label{sym}
\end{equation}
Clearly, there is no analog of the projective generator ${\bf X}_{13}$.
Notice that the first ten
generators are responsible for the obvious symmetry $f\to f+p(x, t)$, here
$p$ is an 
arbitrary cubic polynomial.

I would like to thank A.~V. Aksenov for performing calculations of the
Lie-point symmetries (\ref{sym}).

\section{Acknowledgements}
It is a great pleasure to  thank  M.~A. Akivis, A.~V. Aksenov and Y. Nutku
for their help and clarifying remarks.

\end{document}